\documentclass[preprint,preprintnumbers,amsmath,amssymb]{revtex4-1}
\usepackage{epsfig}

\usepackage{graphicx}
\usepackage{dcolumn}
\usepackage{bm}
\usepackage{chemarr}
\usepackage{multirow}
\usepackage{subfigure}
\usepackage{epstopdf}
\usepackage{arydshln,amsmath,amssymb,float}
\usepackage{color}

\DeclareGraphicsExtensions{.png,.gif,.jpg,.pdf,.bmp}

\begin{document}

\vspace{1.0cm} 

\begin{center}

{\bf Steady-state fluctuations of a genetic feedback loop: an exact solution}

\vspace{0.75cm}

R. Grima$^{1}$, D. R. Schmidt$^{2,3}$, and T. J. Newman$^{4}$

\vspace{0.75cm}

{\it $^{1}$SynthSys Edinburgh, School of Biological Sciences, \\
University of Edinburgh, Edinburgh, EH9 3JR, UK}\\
{\it $^{2}$Mathematical Biosciences Institute, The Ohio State University,\\
Columbus, OH 43210, USA}\\
{\it $^{3}$Computational Biomathematics Laboratory,
Case Western Reserve University, \\ Cleveland, OH 44106, USA}\\
{\it $^{4}$Division of Cell and Developmental Biology, \\
College of Life Sciences, University of Dundee, Dundee, DD1 5EH, UK}
\vspace{0.25cm}
   
\end{center}
  
\begin{center}
{\bf Abstract}
\end{center}
Genetic feedback loops in cells break detailed balance and involve bimolecular reactions; hence exact solutions revealing the nature of the stochastic fluctuations in these loops are lacking. We here consider the master equation for a gene regulatory feedback loop: a gene produces protein which then binds to the promoter of the same gene and regulates its expression. The protein degrades in its free and bound forms. This network breaks detailed balance and involves a single bimolecular reaction step. We provide an exact solution of the steady-state master equation for arbitrary values of the parameters, and present simplified solutions for a number of special cases. The full parametric dependence of the analytical non-equilibrium steady-state probability distribution is verified by direct numerical solution of the master equations. For the case where the degradation rate of bound and free protein is the same, our solution is at variance with a previous claim of an exact solution (Hornos et al, Phys. Rev. E {\bf 72}, 051907 (2005) and subsequent studies). We show explicitly that this is due to an unphysical formulation of the underlying master equation in those studies.

\newpage
 
\section{Introduction}

Biochemical reaction networks underpin the robustness of cells to both internal and external perturbations. Feedback and non-linearities make network behaviour hard to understand, and thus mathematical modelling of the networks can provide useful insights. Copy numbers of gene products, such as proteins, are often relatively small \cite{Ishihama2008,GrimaSchnell2008}, which argues for a careful evaluation of the role of stochasticity. The importance of stochasticity is without doubt in gene expression, given that there are only one or two copies of most genes per cell \cite{Eldar2010}. 
Modelling of the stochastic dynamics of networks is typically more involved than sets of deterministic rate equations. Exact solutions have been obtained for reaction networks obeying detailed balance \cite{Haken,Laurenzi,Cmt} and for those composed of first-order (unimolecular) reactions \cite{Darvey,Gadgil,Heuett,Jahnke,Swain}. However these restrictions are not typical of biochemical processes inside living cells. Detailed balance conditions are characteristic of closed systems of reversible chemical reactions in thermal equilibrium conditions; they only hold for open systems in special cases \cite{Haken}. Living cells are open biochemical systems which actively exchange matter with their surroundings and which possess non-equilibrium steady states, and hence it is clear that the principle of detailed balance will not generally hold for intracellular biochemical systems \cite{Qian2006}. It is also a fact that most systems of interest involve a number of second-order (bimolecular) reactions such as substrate-enzyme interactions, and protein-DNA interactions. 

Here we focus on perhaps the simplest example of a biochemical reaction network which overtly breaks detailed balance and which involves both unimolecular and bimolecular reaction steps. We consider a genetic regulatory network with a feedback loop, namely one in which the product of a gene binds to the promoter of that same gene, and regulates its expression. Furthermore the free and bound protein are assumed to be degraded via proteolysis. Note that while bound protein degradation is not as well known or obvious as free protein degradation, there are mechanisms which could mediate it, e.g. the ubiquitin-proteasome pathway can target and degrade parts of protein complexes \cite{Hochstrasser,Johnson}. Similar feedback mechanisms as discussed above are ubiquitous in biology, appearing in such diverse contexts as metabolism \cite{Selkov1968}, signaling \cite{Goldbeter1975}, somitogenesis \cite{Lewis2003} and circadian clocks \cite{Tyson1999}. Naturally, given its simplicity, special cases of this model have already been the subject of a number of studies. Hornos et al \cite{hornos2005} and subsequent follow up studies \cite{schultz2007,ramos2010,ramos2011} have claimed an exact solution for the case where the rate of bound protein degradation is equal to the rate of free protein degradation. Qian et al \cite{qian2009} have studied the case where the bound protein degradation rate is zero and developed an approximate solution of the master equation in the limits of slow and rapid switching of the gene between the unbound and bound states. A few studies have also considered variations of this simple model by allowing inhibition or activation by protein dimers \cite{Bundschuh2003,Feng2011}. 

In this paper we obtain an exact solution of the master equation for arbitrary free and bound protein degradation rates in non-equilibrium steady state conditions. For the case of equal bound and free protein degradation rates, our solution differs markedly from the exact solution claimed by Hornos et al \cite{hornos2005}; we explicitly show that the difference between the two solutions stems from the fact that the master equations studied in the latter work have no consistent physical interpretation and hence constitute an incorrect description of the biochemical processes at play. 

In the next section we define the model and write down a master equation formulation of the stochastic dynamics. In section III we present the exact solution using the generating function method. In section IV we study three special cases in which detailed balance does not hold and contrast with the case in which detailed balance holds. In section V we directly compare our exact solution with numerical solutions of the master equations and show their correctness. In section VI we present a careful comparison of our exact solution to previous studies. We summarise our results and conclude in section VII.

\section{The model and the master equation}

We consider a single gene and its accompanying promoter region. Self-regulation means that the protein corresponding to the gene can bind to the promoter region and thereby affect the transcription and translation processes. Following previous work, we do not explicitly consider the transcription process, and the intermediate mRNA. Rather we model the process by following only the number of free proteins and the state of the promoter, namely, whether it is bound or unbound. It is important for us to carefully define the processes in order to make completely transparent how these will be encoded into a master equation. By `free proteins' we mean proteins that have been created from transcription/translation of the gene in question, and which are neither bound to the promoter nor degraded. We only allow one protein to be bound to the promoter region at any given time. We do not consider dimerization of free proteins.

We define two conditional probabilities: $P_{0}(n,t)dt$ is the probability that in the time interval $(t,t+\delta t)$ there are $n$ free proteins and the promoter is unbound, and $P_{1}(n,t)dt$ is the probability that in the interval $(t,t+\delta t)$ there are $n$ free proteins and the promoter is bound. In the latter case, there are in fact $n+1$ proteins in the system, $n$ of which are free, and one of which is bound to the promoter. The transcription process will be altered if the promoter is bound, and so the rate of production of proteins will depend on the state of the promoter region. This dependence of production rate on the promoter state breaks detailed balance and makes analytic solution of this problem non-trivial. We define $r_{u}$ and $r_{b}$ to be the production rates of protein given than the promoter region is unbound or bound respectively. We define $k_{f}$ to be the degradation rate of free proteins. We define $k_{b}$ to be the degradation rate of the bound protein. Allowing $k_{b}$ to be non-zero breaks detailed balance even if $r_{u}=r_{b}$ and, again, makes analytic solution non-trivial, but not impossible as we shall see. Lastly, we define $s_{b}$ to be the binding rate per protein to the promoter region, and $s_{u}$ to be unbinding rate from the promoter region. Note, we assume that if the bound protein is degraded it is removed from the system: explicitly the total number of proteins will decrease by one, the total number of free proteins will remain unchanged, and the state of the promoter will change from bound to unbound. These processes and their accompanying rates are schematically illustrated in the following reaction scheme, where $\mathrm{D}_u$, $\mathrm{D}_b$, and $\mathrm{P}$ represent the unbound DNA, the bound DNA, and the free proteins, respectively:
 \begin{align}
&\mathrm{D}_u \xrightarrow{r_u} \mathrm{D}_u + \mathrm{P}, \ \ \mathrm{D}_b \xrightarrow{r_b} \mathrm{D}_b + \mathrm{P}, \nonumber \\
&\quad \quad \quad \quad \quad \mathrm{P} \xrightarrow{k_f} \O, \nonumber \\ 
&\quad \mathrm{D}_b \xrightarrow{k_b} \mathrm{D}_u, \ \ \mathrm{D}_u + \mathrm{P} \xrightleftharpoons[s_u]{s_b} \mathrm{D}_b \ . 
\label{scheme}
\end{align}
Assuming that each process is an independent Poisson process allows us to encode the dynamics using master equations \cite{vankampen_1992}. The master equations for $P_{0}$ and $P_{1}$ have the following forms
\begin{eqnarray}
\nonumber
\frac{d}{dt}P_{0}(n,t) & = & r_{u} \left ( P_{0}(n-1,t) - P_{0}(n,t) \right ) \\ 
\nonumber
& + & k_{f} \left ( (n+1)P_{0}(n+1,t) - nP_{0}(n,t) \right ) \\
\label{master_0}
& + & k_{b}P_{1}(n,t) + s_{u}P_{1}(n-1,t) - s_{b}nP_{0}(n,t) \ , \\ 
\nonumber \\
\nonumber
\frac{d}{dt}P_{1}(n,t) & = & r_{b} \left ( P_{1}(n-1,t) - P_{1}(n,t) \right ) \\ 
\nonumber
& + & k_{f} \left ( (n+1)P_{1}(n+1,t) - nP_{1}(n,t) \right ) \\
\label{master_1}
& - & k_{b}P_{1}(n,t) - s_{u}P_{1}(n,t) + s_{b}(n+1)P_{0}(n+1,t) \ .
\end{eqnarray}

We have formatted these equations to make as clear as possible the relation of the terms to the corresponding molecular processes. Each line of the above equations corresponds to the processes described on the corresponding line of reaction scheme (\ref{scheme}). The first line of each equation refers to processes originating from the gene, i.e., production of the protein. The second line of each equation refers to processes in the cytosol of the cell, i.e., degradation of free proteins. The third line of each equation refers to processes occurring on the promoter, i.e., degradation of the bound protein, and binding and unbinding of individual proteins on the promoter. 

There is no need to write special boundary conditions for these equations, so long as we impose $P_{0}(n,t)=0$ and $P_{1}(n,t)=0$ for $n<0$. Explicitly, if we insert $n=0$ into the above equations we find
\begin{eqnarray}
\label{master_n=0}
\frac{d}{dt}P_{0}(0,t) & = & -r_{u}P_{0}(n,t) + k_{f}P_{0}(1,t) + k_{b}P_{1}(0,t)  \ , \\
\frac{d}{dt}P_{1}(0,t) & = & -r_{b}P_{1}(n,t) + k_{f} P_{1}(1,t) - k_{b}P_{1}(0,t) - s_{u}P_{1}(0,t) + s_{b}P_{0}(1,t) \ ,
\end{eqnarray}
which correctly describe the time evolution of $P_{0}(0,t)$ and $P_{1}(0,t)$. 

In the next section we present the exact solution of these equations in the steady-state using the generating function method.

\section{Exact solution}

It is convenient to work with a dimensionless time variable, and so we choose to scale time by the rate of degradation of free proteins $k_{f}$, i.e., $\tau = k_{f}t$. We define the dimensionless rates
\begin{eqnarray}
\label{drates}
\rho_{u}  =  r_{u}/k_{f}, \ 
\rho_{b}  =  r_{b}/k_{f}, \
\theta = k_{b}/k_{f}, \
\sigma _{u} = s_{u}/k_{f}, \
\sigma _{b}  =  s_{b}/k_{f} \ .
\end{eqnarray}
For future reference we define the convenient parameters 
\begin{equation}
\Sigma_{b} = 1+\sigma _{b}, \ R=\rho_{u}-\rho_{b}\Sigma_{b} \ .
\end{equation}

In dimensionless variables, the master equations (\ref{master_0}) and (\ref{master_1}) take the form
\begin{eqnarray}
\nonumber
\frac{d}{d\tau}P_{0}(n,\tau) & = & \rho_{u} \left ( P_{0}(n-1,\tau) - P_{0}(n,\tau) \right ) \\ 
\nonumber
& + & \left ( (n+1)P_{0}(n+1,\tau) - nP_{0}(n,\tau) \right ) \\
\label{master_dim_1}
& + & \theta P_{1}(n,\tau) + \sigma_{u}P_{1}(n-1,\tau) - \sigma_{b}nP_{0}(n,\tau) \ , \\
\nonumber \\
\nonumber
\frac{d}{d\tau}P_{1}(n,\tau) & = & \rho_{b} \left ( P_{1}(n-1,\tau) - P_{1}(n,\tau) \right ) \\ 
\nonumber
& + & \left ( (n+1)P_{1}(n+1,\tau) - nP_{1}(n,\tau) \right ) \\
\label{master_dim_2}
& - & \theta P_{1}(n,\tau) - \sigma_{u}P_{1}(n,\tau) + \sigma_{b}(n+1)P_{0}(n+1,\tau) \ .
\end{eqnarray}

We will solve these equations using the generating function method \cite{McQuarrie,Gardiner}. We define the generating functions via
\begin{eqnarray}
\label{gf}
G_{0}(z)& = & \sum \limits _{n=0}^{\infty}z^{n} P_{0}(n) \ , \\
G_{1}(z)& = & \sum \limits _{n=0}^{\infty}z^{n} P_{1}(n) \ .
\end{eqnarray}

Henceforth, we will work in the steady-state, and set $dP_{0}(n)/d\tau = 0$ and $dP_{1}(n)/d\tau = 0$. Summing the master equations (\ref{master_dim_1}) and (\ref{master_dim_2}) over $n$ with a weight of $z^{n}$, one finds the coupled pair of first-order differential equations
\begin{align}
\label{gf0}
&\rho_{u}(z-1)G_{0}-(z-1)G_{0}'+(\theta + \sigma_{u}z)G_{1}-\sigma_{b}z G_{0}' = 0 \ , \\
&\rho_{b}(z-1)G_{1}-(z-1)G_{1}'-(\theta + \sigma_{u})G_{1}+\sigma_{b} G_{0}' = 0 \ .
\label{gf1}
\end{align} 

The obvious way to proceed is to write $G_{1}$ in terms of $G_{0}$ and $G_{0}'$ in Eq. (\ref{gf0}) and then substitute into Eq. (\ref{gf1}). However, this leads to a second-order differential equation for $G_{0}$ which is not of the Riemann type, and so cannot be solved in terms of hypergeometric functions.

The less obvious way to proceed is as follows. We differentiate Eq. (\ref{gf0}) to get an equation involving $G_{0}$, $G_{0}'$, $G_{0}''$, $G_{1}$ and $G_{1}'$, and then use (\ref{gf0}) again to eliminate $G_{0}$ in favour of $G_{0}'$ and $G_{1}$. This leads us to an equation involving $G_{0}'$, $G_{0}''$, $G_{1}$ and $G_{1}'$. Finally by means of Eq. (\ref{gf1}), $G_{0}'$ and $G_{0}''$ can be expressed in terms of $G_{1}$ and its derivatives. This leads us to a second-order differential equation for $G_{1}$ which reads
\begin{equation}
\label{gf1_ode}
A(z)G_{1}''+B(z)G_{1}'+C(z)G_{1} = 0 \ ,
\end{equation} 
with
\begin{eqnarray}
\label{coeff_A}
A(z) & = & 1-\Sigma_{b}z \ , \\
\label{coeff_B}
B(z) & = & (\rho_{u}+\rho_{b}\Sigma_{b})z - ((1+\theta)\Sigma _{b} + \sigma_{u} + \rho_{u} + \rho_{b}) \ , \\
\label{coeff_C}
C(z) & = & \rho_{u}(\theta + \sigma _{u} + \rho _{b}) + \rho_{b}\Sigma _{b} - \rho_{u}\rho _{b}z \ . 
\end{eqnarray} 

Since Eq. (\ref{gf1_ode}) has linear coefficients in $z$ it can be transformed to the differential equation for the confluent hypergeometric function. On writing $G_{1}(z) = e^{az}{\tilde G}_{1}(bz+c)$, and substituting into Eq. (\ref{gf1_ode}) one can determine $a$, $b$, and $c$, which provides the following solution 
\begin{equation}
G_{1}(z) = e^{\rho _{b}z}{\tilde G}_{1}(w) \ ,
\end{equation}
where 
\begin{equation}
\label{w_defn}
w = R\frac {(\Sigma_{b} z-1)}{\Sigma _{b}^{2}}\ .
\end{equation}
and ${\tilde G}_{1}(w)$ satisfies Kummer's equation \cite{abramowitz_1972} (i.e., the confluent hypergeometric differential equation)
\begin{equation}
\label{g_tilde_ode}
w{\tilde G}_{1}''+(\beta - w){\tilde G}_{1}'-\alpha {\tilde G}_{1} = 0 \ ,
\end{equation}
with
\begin{equation}
\label{alpha_defn}
\alpha = \theta + \frac {\sigma_{u} (\rho_{u}-\rho_{b})}{R}  \ ,
\end{equation}
and
\begin{equation}
\label{beta_defn}
\beta = 1 + \theta + \frac {1}{\Sigma _{b}} \left ( \sigma_{u} + \rho _{u} - \frac {\rho _{u}}{\Sigma _{b}} \right ) \ .
\end{equation}
Eq. (\ref{g_tilde_ode}) admits two independent solutions, the Kummer function $M(\alpha, \beta, w)$ and the Tricomi function $U(\alpha, \beta, w)$. The latter is inadmissible as a solution for ${\tilde G}_{1}$, as we require that $P_{1}(n) \rightarrow 0$ for $n \rightarrow \infty$ and that the sum over $n$ of $P_{1}(n)$ is finite. Thus, we have the exact solution of the generating function in the form
\begin{equation}
\label{gf1_exact}
G_{1}(z) = A e^{\rho _{b}z} M(\alpha, \beta, w) \ ,
\end{equation}
where $A$ is a normalization constant. Referring to Eq. (\ref{gf1}) we see that knowledge of $G_{1}(z)$ enables us to find an exact expression for $dG_{0}(z)/dz$. Substituting Eq. (\ref{gf1_exact}) into Eq. (\ref{gf1}) and using the transformation properties of the Kummer function, we find
\begin{equation}
\label{g0_diff}
\frac {dG_{0}(z)}{dz} = Ae^{\rho_{b}z} \Biggl [ \frac {\alpha}{(\Sigma_{b}-1)} M(\alpha +1, \beta, w) - \frac{\sigma _{u}\rho _{b}}{R} M(\alpha, \beta, w)
- \frac {R}{\Sigma _{b}^{2}} \frac {\alpha }{\beta }M(\alpha +1,\beta + 1,w) \Biggr ] \ . 
\end{equation}
It is difficult to extract an explicit solution for $G_{0}(z)$ by integrating this expression, and this is consistent with the fact that the second-order differential equation for $G_{0}$ is not of the Riemann form. However, as we shall see, this is not an impediment to finding explicit expressions for $P_{0}(n)$.

We now proceed to obtain the probability distributions $P_0(n)$ and $P_1(n)$. The probability distribution $P_{1}(n)$ can be retrieved from the generating function via
\begin{equation}
\label{invert_gf}
P_{1}(n) = \frac {1}{n!}\frac {d^{n}}{dz^{n}}G_{1}(z) \ \Biggl | _{z=0} \ .
\end{equation}
Substituting Eq. (\ref{gf1_exact}) in the above equation leads us to
\begin{equation}
\label{p1_exact}
P_{1}(n) = \frac {A}{n!} \sum \limits _{m=0}^{n} C_{m}^{n}\rho_{b}^{n-m}\left ( \frac {R}{\Sigma_{b}} \right )^{m} \frac {(\alpha)_{m}}{(\beta)_{m}}M(\alpha+m,\beta + m, w_{0}) \ ,
\end{equation}
where we have defined $w_{0}=w(0)=-R/\Sigma _{b}^2$, the Pochhammer symbol $(a)_{n} = \Gamma (a+n)/\Gamma (a) = a(a+1)\dots (a+n-1)$, with $(a)_{0}=1$, and the combinatorial symbol $C_{m}^{n} = n!/m!(n-m)!$.

Using the analogous expression to Eq. (\ref{invert_gf}) for $G_{0}$ and $P_{0}$, we can obtain $P_{0}(n)$ for $n \ge 1$ by differentiating Eq. (\ref{g0_diff}), with respect to $z$, $(n-1)$ times, dividing by $n!$ and setting $z=0$. With some use of the transformation equations for the Kummer functions we find the compact expression
\begin{eqnarray}
\nonumber
P_{0}(n) = \frac {A}{n!} \sum \limits _{m=0}^{n-1} C_{m}^{n-1}\rho_{b}^{n-1-m}\left ( \frac {R}{\Sigma_{b}} \right )^{m} \frac {(\alpha )_{m}}{(\beta )_{m}} 
& \Biggl \lbrack & \frac {\Sigma _{b}}{(\Sigma _{b}-1)} \left ( m+ \alpha \right ) M(\alpha+m+1,\beta + m, w_{0}) 
\\
\label{p0_exact}
& - & \left ( m+ \alpha + \frac {\sigma _{u}\rho_{b}}{R} \right ) M(\alpha+m,\beta + m, w_{0})  \Biggr \rbrack,
\end{eqnarray}
for $n \ge 1$.
This just leaves $P_{0}(0)$, which can be found directly from the master equation (\ref{master_dim_1}) on setting $n=0$ in the steady-state:
\begin{equation}
\label{p0_0}
\rho _{u} P_{0}(0) = P_{0}(1) + \theta P_{1}(0) \ .
\end{equation}
On substituting the exact forms for $P_{1}(0)$ and $P_{0}(1)$ from Eqs. (\ref{p1_exact}) and (\ref{p0_exact}) respectively, we find
\begin{equation}
\label{p0_0_exact}
P_{0}(0) = A \left \lbrack \frac{\Sigma _{b}}{(\Sigma _{b}-1)} \frac {\alpha }{\rho_{u}} M(\alpha+1,\beta, w_{0}) - \frac {\sigma_{u}}{R} M(\alpha,\beta , w_{0}) \right \rbrack \ .
\end{equation}
Note, this last expression requires $\rho _{u} > 0$ for $P_{0}(0)$ to be finite. In fact, this condition is required for the existence of a non-empty steady-state. If $\rho _{u}=0$ then there is an absorbing state of zero proteins, and the steady-state in that case will correspond to an empty system.

The evaluation of $A$ requires us to normalise by calculating the sum over $n$ of $P_{0}(n)$ and $P_{1}(n)$
\begin{equation}
\label{norm_defn}
\sum \limits _{n=0}^{\infty} (P_{0}(n)+P_{1}(n)) = 1 \ .
\end{equation}
The sum over $P_{1}(n)$ is straightforward, since it is just $G_{1}(1)$. However, the sum over $P_{0}(n)$ cannot be reduced to a simple form as we do not have an explicit expression for $G_{0}(z)$. One can perform the sum over $n$ of $P_{0}(n)$ using the explicit expression (\ref{p0_exact}), but this cannot be reduced beyond definite integrals over Kummer functions whose form is apparently unknown. As such, the normalisation constant $A$ is most easily found by numerically computing the sums over $n$ of the explicit expressions for $P_{0}(n)$ and $P_{1}(n)$.  The probability that there are $n$ proteins in steady-state conditions, $P(n)$, is then given by the sum of $P_{0}(n)$ and $P_{1}(n)$. Ratios of moments such as the Fano factor, i.e. the variance of fluctuations divided by the mean number of molecules, do not depend on $A$ and hence explicit expressions can always be written down for such quantities.

We finish this section by noting that the exact expressions, Eqs. (\ref{p1_exact}), (\ref{p0_exact}) and (\ref{p0_0_exact}), are valid for $R \ne 0$. The singular case $R = 0$ requires special attention and is treated in Appendix A.  

\section{Special cases}
 
In this section, we provide results for four special cases which can be grouped into two classes: (i) the case of detailed balance ($\rho _{u}=\rho _{b}$ and $\theta =0$), in which Poisson statistics are expected to hold, (ii) three cases in which detailed balance does not hold: $\rho _{b}=0$, $\alpha = 0$ and $\alpha = \beta$. In each of these cases one can write down simple expressions for the normalisation constant of the probability distribution and hence one can also obtain explicit expressions for all the moments of the distribution. In what follows we calculate the dependence of the fraction of time for which the promoter is bound on the average number of free proteins. We denote the former quantity by $f_{\rm off}$, which is defined explicitly by 
\begin{equation}
\label{foff_defn}
f_{\rm off} = \sum \limits _{n=0}^{\infty} \ P_{1}(n) \ ,
\end{equation}
and denote the average number of free proteins by $\langle n \rangle$, which is defined explictly by 
\begin{equation}
\label{avn_defn}
\langle n \rangle = \sum \limits _{n=0}^{\infty} \ n (P_{0}(n)+P_{1}(n)) \ .
\end{equation}
We will find that the functional dependence $f_{\rm off}(\langle n \rangle)$ under non-detailed balance conditions can differ markedly from the detailed balance case. 

\subsection{Detailed balance conditions}

The self-regulating gene, described purely in terms of the states representing the number of free proteins, does not generally satisfy detailed balance. This has two causes: (i) the degradation of bound protein, and (ii) the differing rates at which proteins are produced depending on the state of the promoter. Thus, detailed balance can be restored by taking $\theta=0$ {\it and} $\rho _{u}=\rho _{b}$ (see for example \cite{berg_2000} for a general discussion of detailed balance in a gene regulation context). In this case, production of protein is independent of the state of the promoter, and degradation occurs only in the free protein pool. As such, detailed balance holds, and, indeed, the distribution of free protein is trivially Poisson. A well known, yet non-trivial, observation is the simple relationship between $f_{\rm off}$ and $\langle n \rangle$ in this case, which satisfies the Hill function
\cite{Fersht}
\begin{equation}
\label{hill_fn}
f_{\rm off} = \frac {\langle n \rangle}{\langle n \rangle + \frac{s_{u}}{s_{b}}} \ .
\end{equation}

This provides a test of our exact solution which we now confirm. We equate the scaled production rates to each other, defining in the process $\rho \equiv \rho_{u}=\rho _{b}$. We first note from Eq. (\ref{alpha_defn}) that when $\rho_{u}=\rho_{b}$ and $\theta = 0$, then $\alpha = 0 $. The Kummer function $M(\alpha=0, \beta, w)=1$, and so, from Eq. (\ref{gf1_exact}) we have
\begin{equation}
\label{gf1_db}
G_{1}(z) = Ae^{\rho z} \ .
\end{equation}
 
It is straightforward to integrate Eq. (\ref{g0_diff}) for $G_{0}$ in this case, and one finds 
\begin{equation}
\label{gf0_db}
G_{0}(z) = A\frac {\sigma_{u}}{\rho \sigma_{b}} e^{\rho z} + B \ ,
\end{equation}
where $B$ is an integration constant. This constant can be fixed by imposing the condition (\ref{p0_0}) and one finds $B=0$. Normalisation fixes $A$, and one can then retrieve the explicit solutions for the probability distributions:
\begin{eqnarray}
\label{pd0_db}
P_{0}(n) & = & \frac {e^{-\rho}}{\left (1+ \frac{\rho \sigma_{b}}{\sigma_{u}} \right )} \frac {\rho^{n}}{n!} \ , \\ 
\label{pd1_db}
P_{1}(n) & = & \frac {e^{-\rho}}{\left (1+ \frac{\sigma_{u}}{\rho \sigma_{b}} \right )} \frac {\rho^{n}}{n!} \ . 
\end{eqnarray}
Note, the distribution of free proteins, $P_{0}(n)+P_{1}(n)=e^{-\rho}\rho ^{n}/n!$: a Poisson distribution as anticipated. Summing these distributions over $n$ with a weight of $n$, the average number of proteins is easily found to be $\langle n \rangle = \rho$. Now the fraction of time for which the promoter is bound, $f_{\rm off}$, is equal to the probability that the gene is `off', which is obtained by summing Eq. (\ref{pd1_db}) over $n$. Expressing the latter in terms of $\langle n \rangle$ and reverting to unscaled rate parameters we obtain Eq. (\ref{hill_fn}), as required.

We finish this section by noting that the deterministic model of the genetic feedback loop predicts that $f_{\rm off}$ has a Hill function dependence on $\langle n \rangle$ for all parameter values (see Appendix B). Hence for the detailed balance case, the predictions of the stochastic and deterministic models are in agreement.

\subsection{Non-detailed balance conditions}

\subsubsection{The case $\rho _{b}=0$}

This is the case of strong transcriptional repression since the production of the gene in the bound state is zero. In this case, we see from Eq. (\ref{g0_diff}) that the expression for $dG_{0}(z)/dz$ is a sum of Kummer functions, which can be directly integrated. On doing so, and, again, utilising the transformation formulae for Kummer functions, one finds the compact result
\begin{equation}
\label{g0_case1}
G_{0}(z) = \frac {A}{\rho _{u}(\Sigma _{b}-1)} \Biggl [ (\theta + \sigma _{u}) \Sigma _{b} M(\alpha +1, \beta, w) - \sigma _{u}(\Sigma _{b}-1) M(\alpha, \beta, w) \Biggr ] \ .  
\end{equation}
In principle, an unknown constant $B$ should be added to this expression; however, one can fix this constant by utilising the relation (\ref{p0_0}) and one finds that $B=0$. Directly from Eq. (\ref{gf1_exact}), we have
\begin{equation}
\label{gf1_case1}
G_{1}(z) = A M(\alpha, \beta, w) \ ,
\end{equation}
where we have the simpler expressions $\alpha = \theta + \sigma _{u}$ and $w(z)=\rho _{u}(\Sigma _{b}z-1)/\Sigma _{b}^{2}$; the parameter $\beta $ is still given by Eq. (\ref{beta_defn}). 

It is now straightforward to determine the normalisation constant $A$ by imposing condition (\ref{norm_defn}), which is equivalent to $G_{0}(1)+G_{1}(1)=1$. One finds
\begin{equation}
\label{norm_exact}
A=\rho _{u} (\Sigma _{b}-1){\tilde A} \ ,
\end{equation}
where
\begin{equation}
\label{norm_exact_2}
{\tilde A} = \Bigl [ (\theta + \sigma _{u}) \Sigma _{b} M(\alpha +1, \beta, w_{1}) + (\rho _{u} - \sigma _{u})(\Sigma _{b}-1) M(\alpha, \beta, w_{1}) 
\Bigr ] ^{-1} \ ,
\end{equation}
and $w_{1} = w(1) = \rho _{u}(\Sigma _{b}-1)/\Sigma _{b}^{2} $.  
 
The explicit forms for the probability distributions can be obtained directly from the general formulae (\ref{p1_exact}), (\ref{p0_exact}), and (\ref{p0_0_exact}), or from direct evaluation from the explicit generating functions (\ref{g0_case1}) and (\ref{gf1_case1}). In either case, one obtains
\begin{equation}
\label{p0_exact_case1}
P_{0}(n) = \frac {{\tilde A}}{n!} \left ( \frac {\rho _{u}}{\Sigma_{b}} \right )^{n} \frac {(\alpha)_{n}}{(\beta)_{n}} \Bigl [ (\alpha + n)\Sigma _{b} 
M(\alpha+n+1,\beta + n, w_{0}) - \sigma _{u}(\Sigma _{b} -1) M(\alpha+n,\beta + n, w_{0}) \Bigr ] \ ,
\end{equation}
and
\begin{equation}
\label{p1_exact_case1}
P_{1}(n) = \frac {A}{n!} \left ( \frac {\rho _{u}}{\Sigma_{b}} \right )^{n} \frac {(\alpha)_{n}}{(\beta)_{n}}M(\alpha+n,\beta + n, w_{0}) \ ,
\end{equation}
where in this case $w_{0}=w(0)=-\rho _{u}/\Sigma _{b}^{2}$. Note that Eq. (\ref{p0_exact_case1}) for $P_{0}(n)$ is valid for all $n \ge 0$. 

The fraction of time for which the promoter is bound, $f_{\rm off}$, is given by $G_{1}(1)$, and has the form
\begin{equation}
\label{frac_off}
f_{\rm off} = \rho_u (\Sigma _{b}-1){\tilde A}  M(\alpha, \beta, w_{1}) \ . 
\end{equation}

The mean number of free proteins, regardless of the state of the gene is given by $G_{0}'(1)+G_{1}'(1)$ and has the form
\begin{equation}
\label{av_n}
\langle n \rangle =  \rho _{u} \alpha {\tilde A} \Bigl [ \Sigma _{b} M(\alpha +1, \beta, w_{1}) - (\Sigma _{b}-1)M(\alpha, \beta, w_{1}) 
\Bigr ] \ . 
\end{equation}

Unlike the detailed balance case, it does not seem generally possible to write $f_{\rm off}$ as a function of $\langle n \rangle $. The behavior for small and large $\langle n \rangle$ can however be easily deduced. It is clear that in the limit of large $\langle n \rangle$, $f_{\rm off}$ approaches one. The small $\langle n \rangle$ behavior can be inferred by a series expansion of Eqs. (\ref{frac_off}) and (\ref{av_n}) in powers of $\rho_u$
\begin{align}
f_{\rm off} &= \frac{\sigma_b \rho_u}{\theta \Sigma_b + \sigma_u} + O(\rho_u^2), \\
\langle n \rangle &= \frac{(\theta + \sigma_u) \rho_u}{\theta \Sigma_b + \sigma_u} + O(\rho_u^2).
\end{align}
From these expressions, one can deduce that
\begin{equation}
f_{\rm off} \simeq \frac{\sigma_b \langle n \rangle}{\theta + \sigma_u}.
\label{smalln}
\end{equation}
The intermediate $\langle n \rangle$ behavior is however unknown. The dependence of $f_{\rm off}$ with $\langle n \rangle$ is most easily explored by numerically evaluating Eqs. (\ref{frac_off}) and (\ref{av_n}) for various values of $\rho_u$. In Fig. 1 we show two plots generated in this manner, one for $\sigma_b$ small and the other for $\sigma_b$ large, both with $\theta = 0$. In each case we compare with the Hill function 
\begin{equation}
f_{\rm off}^* = \frac{\langle n \rangle}{\langle n \rangle + \frac{\theta + \sigma_u}{\sigma_b}}.
\label{hillapprox}
\end{equation}
Note that this function's small and large $\langle n \rangle$ dependence are the same as those of the exact solution. Furthermore this function is the prediction of the deterministic model (see Appendix B).
From Fig. 1, we see that the Hill function is a good approximation to the actual function for small $\sigma_b$; in the opposite limit of large $\sigma_b$, the two functions are considerably different for intermediate $\langle n \rangle$. In this case the exact solution shows a piecewise linear form, with the linear dependence of $f_{\rm off}$ approximately holding until the function `breaks' at its threshold value of unity. 

As a last comment in this subsection, defining by $\langle n \rangle _{0}$ the average of $n$ conditioned on the promoter being unbound, i.e.
\begin{equation}
\label{av_n_0_def}
\langle n \rangle _{0} =  \frac {\sum \limits _{n=0}^{\infty} n \ P_{0}(n)} {\sum \limits _{n=0}^{\infty} P_{0}(n)} \ , 
\end{equation}
we find the simple form
\begin{equation}
\label{av_n_0}
\langle n \rangle _{0} =  \frac {\rho _{u} {\tilde A} (\theta + \sigma _{u}) M(\alpha, \beta, w_{1})}{(1-f_{\rm off})} \ . 
\end{equation}
Comparing this with Eq.(\ref{frac_off}) we have
\begin{equation}
\label{av_n_0p}
\langle n \rangle _{0} =  \frac {(\theta + \sigma _{u})}{(\Sigma _{b}-1)} \  \frac {f_{\rm off}} {(1-f_{\rm off})} \ , 
\end{equation}
which can be inverted to obtain the curious result
\begin{equation}
f_{\rm off} = \frac { \langle n \rangle _{0} } { \langle n \rangle _{0} + \frac {(\theta + \sigma _{u})}{\sigma _{b} } }  \ .
\label{f_off_av_n_0}
\end{equation}
Thus, an equation resembling the Hill function is found, but the average of $n$ is replaced by the average of $n$ conditioned on the promoter being unbound (i.e. the gene being switched on). In fact, one can show that this relationship holds quite generally (i.e. for arbitrary values of $\rho_{b}$) by summing the master equation (\ref{master_dim_1}) for $P_{0}(n)$ over $n$ and using the definitions of $f_{\rm off}$ and $\langle n \rangle _{0}$. 
 
\subsubsection{The case $\alpha =0$}
 
We saw in the subsection on detailed balance conditions that when $\alpha = 0$ the Kummer function is a constant, and $G_{1}$ then simplifies to an exponential function, giving Poisson statistics for $P_{1}(n)$ (and also for $P_{0}(n)$ in that case). The cases considered in this subsection and the next are two further cases in which the exact solution for $G_{1}$ reduces to a pure exponential function. Note, we have previously defined $R=\rho _{u}-\rho _{b}\Sigma _{b}$. We will not be using $R$ for this case and the next one, as it is helpful to see the role of $\rho _{u}$ and $\rho _{b}$ explicitly.

In setting $\alpha =0$, but without imposing the two conditions for detailed balance, we have 
\begin{equation}
\label{gf1_op}
G_{1}(z) = Ae^{\rho _{b}z} \ ,
\end{equation}
and Poisson statistics for $P_{1}(n)$. The condition $\alpha=0$ can be written
\begin{equation}
\label{alpha=0}
\theta = \frac {\sigma _{u}(\rho _{u}-\rho _{b} )}{(\rho _{b}\Sigma _{b}-\rho _{u})} \ .
\end{equation}
To hold, this relation requires a quite severe constraint on the range of $\rho _{u}$, namely $\rho_{b} \le \rho _{u} < \rho _{b}\Sigma _{b}$. 
Given the form of $G_{1}$ in Eq. (\ref{gf1_op}), one can use Eq. (\ref{g0_diff}) to retrieve the simple exponential form for $G_{0}$ and the normalisation condition and Eq. (\ref{p0_0}) to fix the two arbitrary constants that arise. After some algebra one finds
\begin{eqnarray}
\label{pd0_op}
P_{0}(n) & = & \frac {e^{-\rho_{b}}}{\left (1+ \frac{(\rho _{b} \Sigma_{b}-\rho_{u})}{\sigma_{u}} \right )} \frac {\rho_{b}^{n}}{n!} \ , \\ 
\label{pd1_op}
P_{1}(n) & = & \frac {e^{-\rho_{b}}}{\left (1+ \frac{\sigma_{u}}{(\rho _{b}\Sigma_{b}-\rho_{u})} \right )} \frac {\rho_{b}^{n}}{n!} \ .
\end{eqnarray}
The mean number of free proteins has the particularly simple form $\langle n \rangle = \rho _{b}=r_{b}/k_{f}$.
 
In this and the next subsection the special cases impose non-trivial relationships between parameters (see Eqs (\ref{alpha=0}) and (\ref{alpha=beta})). As such, these solutions are valid on hypersurfaces in parameter space. Because of this non-trivial relationship between parameters, one cannot change $\langle n \rangle$ through variation of a single parameter, and as such the functional relationship $f_{\rm off} (\langle n \rangle)$ is of limited experimental interest, and thus we do not report such results here.

\subsubsection{The case $\alpha=\beta$}

In setting $\alpha=\beta$ we can take advantage of the fact that $M(\alpha, \alpha, w)=e^{w}$ \cite{abramowitz_1972}. Thus, using Eqs. (\ref{w_defn}) and (\ref{gf1_exact}) we have
\begin{equation}
\label{gf1_op2}
G_{1}(z) = A'e^{\rho _{u}z/\Sigma _{b}} \ ,
\end{equation}
and, consequently, Poisson statistics for $P_{1}(n)$. Using Eqs. (\ref{alpha_defn}) and (\ref{beta_defn}), the condition $\alpha = \beta$ translates to fixing $\sigma _{u}$ in terms of $\rho _{u},\rho_{b}$, and $\Sigma _{b}$, independent of $\theta$, as follows
\begin{equation}
\label{alpha=beta}
\sigma_{u} = \frac {(\rho _{u}-\rho _{b}\Sigma _{b})(\Sigma _{b}^{2}+\rho _{u}(\Sigma _{b}-1))}{\rho _{u} \Sigma _{b}(\Sigma _{b}-1)} \ .
\end{equation}
Note, this condition requires $\rho _{u} > \rho_{b}\Sigma _{b}$, and so this condition has no overlap with the condition $\alpha=0$ studied above. The implicit reason for this stems from the definitions of $\alpha$ and $\beta$: while the former can take a value of zero, the latter is always greater than 1. 

One can now substitute the expression for $G_{1}(z)$ into Eq. (\ref{g0_diff}), and find the explicit form for $G_{0}(z)$
\begin{equation}
\label{gf0_op2}
G_{0}(z) = \frac {A'}{(\Sigma _{b}-1)} \left \lbrack \left ( \frac{\rho _{u}-\rho _{b}\Sigma _{b}}{\rho _{u}} \right )z + \frac{\Sigma _{b}}{\rho _{u}} \left ( \theta + \frac {(\rho _{u}-\rho _{b}\Sigma _{b})}{\rho _{u}(\Sigma _{b}-1)} \right )  \right \rbrack e^{\rho _{u}z/\Sigma _{b}} \ .
\end{equation}
Note, the arbitrary constant which arises from integrating Eq. (\ref{g0_diff}) is found to be zero on application of the condition (\ref{p0_0}). Given the $z$ dependence of the prefactor to the exponential, $P_{0}(n)$ will not have a purely Poisson form.

After some algebra, one finds the explicit forms for the probability distributions:
\begin{eqnarray}
\label{pd0_op2}
P_{0}(n) & = & \frac {A'}{(\Sigma _{b}-1)} \left \lbrack \left ( \frac{\rho _{u}-\rho _{b}\Sigma _{b}}{\rho _{u}} \right )  \frac {(\rho_{u}/\Sigma _{b})^{n-1}}{(n-1)!} + \frac{\Sigma _{b}}{\rho _{u}} \left ( \theta + \frac {(\rho _{u}-\rho _{b}\Sigma _{b})}{\rho _{u}(\Sigma _{b}-1)} \right )  \frac {(\rho_{u}/\Sigma _{b})^{n}}{n!} \right \rbrack  \ , \\ 
\label{pd1_op2}
P_{1}(n) & = & A' \frac {(\rho_{u}/\Sigma _{b})^{n}}{n!} \ ,
\end{eqnarray}
where
\begin{equation}
\label{norm_op2}
A'=\frac 
{\rho _{u}^{2}(\Sigma _{b}-1)^{2}e^{-\rho _{u}/\Sigma _{b}} }
{\Sigma _{b} \left [ (\rho _{u}-\rho_{b}\Sigma _{b}) + \rho _{u}(\Sigma _{b}-1)(\theta +\rho _{u} -\rho _{b})  \right ] } \ .
\end{equation}
Note, Eq. (\ref{pd0_op2}) holds for $n=0$ with the understanding that $1/(-1)!=0$.

Using these distributions, we find
\begin{equation}
\label{avn_op2}
\langle n \rangle = \rho _{u} \left [ \frac
{(\rho _{u}-\rho_{b}\Sigma _{b})\Sigma _{b} + \rho _{u}(\Sigma _{b}-1)(\theta +\rho _{u} -\rho _{b}) }
{(\rho _{u}-\rho_{b}\Sigma _{b}) + \rho _{u}(\Sigma _{b}-1)(\theta +\rho _{u} -\rho _{b}) } \right ]
\ .
\end{equation}
Note that $\langle n \rangle > \rho _{u}$, and hence $\langle n \rangle > \rho _{u}/\Sigma _{b}$ which is the parameter in the Poisson-like distributions. 
 
\section{Numerical validation of the exact solution}

In this section we numerically solve the master equations, Eqs. (\ref{master_dim_1})-(\ref{master_dim_2}) and compare with the exact solutions obtained in Section III. $2N$ difference equations are generated by substituting $n =0, 1, 2, ..., N-1$ in Eqs. (\ref{master_dim_1})-(\ref{master_dim_2}) with the time derivative set to zero. The boundary conditions are set to $P_0(-1) = P_1(-1)=P_0(N)=P_1(N)=0$, taking into account also the absence of probability flux into $P_{0}(N)$ and $P_{1}(N)$. This set of difference equations is solved simultaneously for $P_0(0), P_1(0), P_0(1), P_1(1), ..., P_0(N-1), P_1(N-1)$. Note that the exact solution corresponds to $N$ equals positive infinity. Of course practically we are only interested in obtaining the probability distribution solution to some desired accuracy and hence it is sufficient to solve the equations for a large enough positive integer $N$. This should be chosen large enough such that the probability distribution solution $P_0 (n) + P_1 (n)$ smoothly decays to zero as $n$ approaches $N$.  

We use the latter method with $N = 500$ to obtain the dependence of the steady-state probability distribution solution of the master equations, Eqs. (\ref{master_dim_1})-(\ref{master_dim_2}), on the five non-dimensional parameters $\theta$, $\sigma_b$, $\sigma_u$, $\rho_b$ and $\rho_u$. The same is obtained by means of the analytical solutions given by Eqs. (\ref{p1_exact}), (\ref{p0_exact}) and (\ref{p0_0_exact}). The results from the two methods are compared in Fig. 2 where the open circles show the numerics and the crosses show the analytical solution. Note that in all cases, $P(n)$ goes to zero for $n$ much less than 500 and hence artificial boundary effects due to finite $N$ should be negligible; indeed we verified that the probability distribution solution obtained with $N = 1000$ is indistinguishable from the one obtained with $N = 500$. Note that the numerics and the analytical solution are in perfect agreement which indeed verifies the correctness of the main result of this paper, namely the exact solution given by Eqs. (\ref{p1_exact}), (\ref{p0_exact}) and (\ref{p0_0_exact}). 

It is interesting that in (a), (b), as we gradually increase $\theta$ and $\sigma_b$ respectively, we observe a transition from a bimodal to a unimodal probability distribution while in (c) we see the reverse transition as the parameter $\rho_u$ is increased. The bimodal character of the distribution is particularly interesting since the deterministic model of the genetic feedback loop does not exhibit bistability (see Appendix B). The mechanism behind the origin of bimodality and the transition from bimodal to unimodal behavior can be inferred as follows. In cases (a)-(c), the peaks of the bimodal distribution occur at $n \simeq \rho_u$ and at $n \simeq \rho_b$ while the single peak of the unimodal distribution occurs at one of these two (depending on parameter values). Now the processes $\mathrm{D}_u \xrightarrow{r_u} \mathrm{D}_u + \mathrm{P},$ and $\mathrm{P} \xrightarrow{k_f} \O$ considered by themselves lead to a Poisson distribution for the number of protein molecules with peak at $n \simeq \rho_u$ in steady-state conditions; similarly a peak at $n \simeq \rho_b$ can be associated with the processes $\mathrm{D}_b \xrightarrow{r_b} \mathrm{D}_b + \mathrm{P},$ and $\mathrm{P} \xrightarrow{k_f} \O$. Hence it is clear that bimodality occurs whenever the gene switches slowly between its bound and unbound states which leads to a switch between the two sets of reactions discussed above. Inspection of the reaction mechanism Eq. (\ref{scheme}) shows that the switching rates increase with $\theta$, $\sigma_b$ and $\sigma_u$. Hence if for some parameter set we have bimodality, increasing any one of the aforementioned three parameters will lead to a switch from bimodal to unimodal behavior; these are cases (a) and (b). If we have unimodality then the distribution will of course stay unimodal upon variation of one of the three parameters; this is case (d). Slow transitions between unbound and bound states are necessary but not sufficient to induce bimodality; the protein production rates of the two gene states must be sufficiently different such that the peaks of the Poisson distributions associated with each state are well separated; this is case (c). Cases (a)-(d) are ones in which the genetic feedback loop is negative, i.e. $\rho_b \le \rho_u$. In contrast cases (e) and (f) are for a positive feedback loop, i.e. $\rho_b > \rho_u$. As for the negative feedback case, both unimodal (case (e)) and bimodal behaviors (case (f)) are possible and their existence can be understood by the same switching mechanism elucidated above. For example the case $\theta = 0$ in (f) is bimodal with peaks at $n \simeq \rho_u$ and $n \simeq \rho_b$ indicating slow switching between the steady-states of the bound and unbound genes. Increasing $\theta$ leads to an increase in the switching rate from the bound to the unbound states explaining the increase in the size of the peak associated with the unbound state and the corresponding decrease of the peak size associated with the bound state. 
 
\section{Critique of a previous ``exact'' solution}

In this section we make a careful and explicit comparison of our master equations with those of Hornos et al \cite{hornos2005} in the original paper claiming an exact solution to the problem of a self-regulating gene with the condition $k_{f}=k_{b}$ (i.e. $\theta =1$). In that work, the variable $n$ of the probability distributions represents the {\it total number} of proteins in the system, that is the number of free proteins plus (when the promoter is bound) the bound protein.

The following key provides the translation between the notation we have used in section II and the notation of Hornos et al. 
\begin{eqnarray}
P_{0}(n,t) & \longleftrightarrow & \alpha _{n}(t) \\ 
P_{1}(n,t) & \longleftrightarrow & \beta _{n+1}(t) \\ 
r_{u} & \longleftrightarrow & g_{\alpha} \\
r_{b} & \longleftrightarrow & g_{\beta} \\
k_{f} & \longleftrightarrow & k_{f} \\
k_{b} & \longleftrightarrow & k_{b} \\
s_{u} & \longleftrightarrow & f \\
s_{b} & \longleftrightarrow & h
\end{eqnarray}
Note, in their original work, Hornos et al assume from the outset that the rates of degradation of free and bound protein are the same, and use the symbol $k$ for this degradation rate. Here, for clarity in the handling of these two different processes, we allow the rates to be different, and use the symbols $k_{f}$ and $k_{b}$ consistent with the notation used in section II, setting them equal eventually.

With the key given above, the correct master equations (\ref{master_0}) and (\ref{master_1}) take the form
\begin{eqnarray}
\nonumber
\frac{d}{dt}\alpha _{n}(t) & = & g_{\alpha} \left ( \alpha _{n-1} - \alpha _{n} \right ) \\ 
\nonumber
& + & k_{f} \left ( (n+1)\alpha_{n+1} - n\alpha _{n} \right ) \\
\label{master_0_h}
& + & k_{b}\beta _{n+1} + f \beta_{n} - hn\alpha _{n} \ , \\ 
\nonumber \\
\nonumber
\frac{d}{dt}\beta _{n} & = & g_{\beta} \left ( \beta _{n-1} - \beta _{n} \right ) \\ 
\nonumber
& + & k_{f} \left ( n\beta _{n+1}  - (n-1)\beta _{n} \right ) \\
\label{master_1_h}
& - & k_{b}\beta _{n} - f\beta _{n} + hn \alpha _{n} \ .
\end{eqnarray}
The first equation is valid for $n \ge 0$ while the second equation is valid for $n \ge 1$. The boundary conditions $P_{0}(-1,t)=0$ and $P_{1}(-1,t)=0$ imply the new boundary conditions $\alpha_{-1}=\beta_{0}=0$. 

On setting the free and bound degradation rates to be equal $k_{f}=k_{b}=k$, the master equations above take the form
\begin{eqnarray}
\nonumber
\frac{d}{dt}\alpha _{n}(t) & = & g_{\alpha} \left ( \alpha _{n-1} - \alpha _{n} \right ) \\ 
\nonumber
& + & k \left ( (n+1)\alpha_{n+1} - n\alpha _{n} \right ) \\
\label{master_0_h_eq}
& + & k\beta _{n+1} + f \beta_{n} - hn\alpha _{n} \ , \\ 
\nonumber \\
\nonumber
\frac{d}{dt}\beta _{n} & = & g_{\beta} \left ( \beta _{n-1} - \beta _{n} \right ) \\ 
\nonumber
& + & k \left ( n\beta _{n+1}  - (n-1)\beta _{n} \right ) \\
\label{master_1_h_eq}
& - & k\beta _{n} - f\beta _{n} + hn \alpha _{n} \ .
\end{eqnarray}

We are now in a position to directly compare these master equations with those of Hornos et al. Equations (1) and (2) of that paper read (for $n>0$)
\begin{eqnarray}
\nonumber
\frac{d}{dt}\alpha _{n}(t) & = & g_{\alpha} \left ( \alpha _{n-1} - \alpha _{n} \right ) \\ 
\nonumber
& + & k \left ( (n+1)\alpha_{n+1} - n\alpha _{n} \right ) \\
\label{master_0_h_orig}
& + & f \beta_{n} - hn\alpha _{n} \ , \\ 
\nonumber \\
\nonumber
\frac{d}{dt}\beta _{n} & = & g_{\beta} \left ( \beta _{n-1} - \beta _{n} \right ) \\ 
\nonumber
& + & k \left ( n\beta _{n+1}  - (n-1)\beta _{n} \right ) \\
\label{master_1_h_orig}
& + & k \left (\beta _{n+1} - \beta _{n} \right ) - f\beta _{n} + hn \alpha _{n} \ .
\end{eqnarray}
Obviously, in the second and third lines of Eq. (\ref{master_1_h_orig}) one can combine the terms with a prefactor of $k$ to give $k((n+1)\beta _{n+1} - n\beta _{n})$, but we have separated these terms to emphasise the comparison to the correct equation (\ref{master_1_h_eq}) in which, as stressed in section II, each line of the equation corresponds to processes occurring on the gene, the cytosol, and the promoter respectively.
In order to conserve probability, Hornos et al are obliged to write a separate equation for $n=0$ which reads
\begin{equation}
\label{master_0_h_0}
\frac{d}{dt}\alpha _{0}(t) = -g_{\alpha} \alpha _{0} + k(\alpha _{1}+\beta _{1}) \ , 
\end{equation}
which is consistent with Eq. (\ref{master_0_h_eq}) for $n=0$ and using the boundary condition $\alpha_{-1}=0$. They also enforce the boundary condition $\beta_0 = 0$. The same master equations, Eqs. (\ref{master_0_h_orig}) and (\ref{master_1_h_orig}), appear also in several follow up papers \cite{schultz2007,ramos2010,ramos2011}. 

Let us denote the master equations (\ref{master_0_h_eq}) and (\ref{master_1_h_eq}), as ME(GSN), and the master equations (\ref{master_0_h_orig}) and (\ref{master_1_h_orig}) as ME(H-ET-AL). Clearly ME(GSN) and ME(H-ET-AL) are different, and given they purport to describe the same process, they cannot both be correct. The difference between the two sets of equations is in how degradation of the bound protein is described. In our master equations ME(GSN), which were derived from Eqs. (\ref{master_0}) and (\ref{master_1}) by use of the key, if the bound protein is degraded, the system returns to the unbound state, i.e., that process connects the two conditional probabilities $\alpha _{n}$ and $\beta _{n+1}$. In ME(H-ET-AL), degradation of the bound protein keeps the system in the bound state. How is this possible? Literally, the ME(H-ET-AL) master equations are describing bound protein degradation as the following composite of processes: degradation of the bound protein, with instantaneous rebinding of a protein to the promoter from the free protein pool, such that the state $\beta _{n+1}$ transitions to the state $\beta _{n}$. This composite of processes occurs with a constant rate $k$, independent of the number of free proteins. Clearly this composite of processes is not the intended instantiation of the self-regulating gene, and furthermore, it is difficult to imagine any physical process that could operate in this manner. One could concoct an imaginary process using a Maxwell Demon, a tiny animated figure, who sits by the promoter with a free protein always to hand, to instantaneously latch onto the promoter should the bound protein be degraded, but such constructions belong to philosophical discussions of irreversibility, and are not relevant to the biological question at hand.

One might argue that the manner in which the bound protein degradation is handled has little bearing on the form of the distributions. To test this, we used the method expounded in Sec. V to numerically solve the two master equations, ME(GSN) and ME(H-ET-AL), in steady-state conditions for four different sets of parameter values. The results are shown in Fig. 3, where the open circles show the predictions of ME(GSN) while the solid lines show the prediction of ME(H-ET-AL). The parameters $g_{\alpha}$, $g_{\beta}$, and $k$ are, in all cases, equal to $80.0$, $0.0$, and $1.0$ respectively, while $h = 0.001\Lambda$ and $f = 0.1\Lambda$ where $\Lambda$ takes the values 0.01 in Fig. 3(a), 1 in Fig. 3(b), 10 in Fig. 3(c) and 100 in Fig. 3(d). Note that as $\Lambda$ is varied from 0.01 to 100, the ME(H-ET-AL) predicts a transition from unimodal to bimodal probability distribution and back to unimodal distribution. However no such transitions are seen in ME(GSN). Note that the case $\Lambda = 1$ in Fig. 3(b) corresponds to the exact set of parameters used in the case $\omega = 0.1$ in Fig. 1 of the Hornos et al paper \cite{hornos2005}. The good agreement between ME(GSN) and ME(H-ET-AL) for cases (a) and (d) can be explained as follows. In case (a), the rate of protein binding to the promoter is very small, the gene is most of the time in the unbound state and hence bound protein degradation rarely occurs. In case (d), the gene switches between the unbound and unbound states very rapidly but its decay to the unbound state occurs primarily via the reaction $\mathrm{D}_b \xrightarrow{} \mathrm{D}_u + \mathrm{P}$ and only occasionally via the reaction $\mathrm{D}_b \xrightarrow{} \mathrm{D}_u$. In both cases, bound protein degradation is a rare event compared to the other molecular processes and hence it follows that the form of the probability distribution is practically insensitive to whether bound protein degradation is correctly or incorrectly described. In cases (b) and (c), bound protein degradation events occur at a frequency comparable with that of other molecular processes, and hence its correct description using the ME(GSN) becomes crucial to obtaining the correct steady-state probability distribution. The discussion on the role of $\theta $, particularly when it approaches unity, explains why the bimodal distribution is not to be expected in a correct formulation of this problem.

It is worth mentioning that a cursory comparison of our results with those of Hornos et al indicates some similarities, which are in some cases superficial. For example, Hornos et al find Kummer function solutions for the generating functions. Note, in the solution of ME(H-ET-AL), it is the generating function conditioned on the unbound promoter ($G_{0}$ in our notation) which is expressed in terms of a single Kummer function, while the generating function conditioned on the bound promoter ($G_{1}$ in our notation) is expressed in terms of a sum of Kummer functions. In our exact solution of the correct master equation ME(GSN) we find that $G_{1}$ is the generating function which has the simple form in terms of a single Kummer function, as shown in Eq.(\ref{gf1_exact}), while $G_{0}$ involves integrals of Kummer functions, and does not have a simple closed form expression. We also note that Figure 2 of Hornos et al shows a plot of $f_{\rm off}$ versus $\langle n \rangle$,  which indeed continuously varies from the Hill form to a linear piece-wise form as $\sigma _{b}$ is decreased, similar to our result shown in Figure 1. In the caption of Figure 2 of their paper, the curious relation between $f_{\rm off}$ and $\langle n \rangle _{0}$ is noted, similar to our Eq. (\ref{f_off_av_n_0}),  although the combination $(\theta + \sigma _{u})/\sigma _{b}$ in our equation is replaced by $\sigma _{u}/\sigma _{b}$ in theirs, indicating again the incorrect handling of protein degradation in that work. 

Qian et al \cite{qian2009} studied the case where the bound protein does not degrade (i.e. $\theta = 0$). The coupled master equations, Eqs. 17(a)-17(b) in their paper,  are the same as our master equations, Eqs. (\ref{master_0_h})-(\ref{master_1_h}), with $k_b=0$. Hence their master equations are correct for this special case. However they do not solve these equations exactly. Rather they derive an approximative solution in the limit that the gene switches between the $\mathrm{D}_u$ and $\mathrm{D}_b$ states very rapidly and in the limit that the switching occurs very slowly. In the latter limit, i.e., the limit of small protein binding and unbinding rates to the promoter, they show that the probability distribution is bimodal if the production rates of the gene in the bound and unbound states are sufficiently different. This is confirmed by our exact solution, see the case $\theta = 0$ in Fig. 2 (a).  We find that this bimodal behaviour disappears when $\theta$ is increased from 0 to 1 (see Fig. 2 (b) and (c)).

\section{Conclusions}

In this paper we have presented an exact solution for the simplest model of a self-regulating gene. Our solution is valid for an arbitrary degradation rate of the bound protein, which we have denoted throughout the paper by the parameter $\theta$ in dimensionless units. The explicit solution for the probability distributions for the number of free proteins, conditioned on the gene's promoter being bound or unbound, are given by Eqs. (\ref{p1_exact}) and (\ref{p0_exact}) respectively. It is interesting that an exact solution for this general case can be found given that the model breaks detailed balance and includes a bimolecular reaction step, features not typical of the exact solutions reported in the literature \cite{Haken,Laurenzi,Darvey,Gadgil,Heuett,Swain,Cmt}. In particular, to the best of our knowledge, our exact solution is a first for a gene regulatory network with a feedback loop. As we have shown, the previous exact solution claimed by Hornos et al \cite{hornos2005} for the special case in which bound and free protein degradation are equal is incorrect because it is based on master equations which possess no coherent physical interpretation. The only other exact solution known in the context of stochastic gene expression is that derived by Shahrezaei and Swain for a model of gene expression involving first-order processes describing transcription, translation, protein degradation and mRNA degradation but no feedback loop \cite{Swain}. 

We anticipate that our exact solution will be useful to explore the dependence of common experimental measures of noise intensity, e.g., the coefficient of variation \cite{Bar-even} and Fano factors \cite{Taniguchi}, on various parameter values and on the nature of the feedback loop (repressing or activating). This may lead to insights into the mechanisms used by cells to regulate fluctuations in the protein concentrations, a topic of intense current research \cite{Raser2005,Rao2002}. Other interesting avenues of research, which we have briefly touched upon in this paper, are the investigation of the transition from unimodal to bimodal protein distributions, and of deviations from the Hill function describing activation or repression of the gene. Our exact solution will enable a more thorough analysis of these topics which could previously only be investigated by means of approximation methods in some restricted parameter regimes.

For chemical systems involving bimolecular reactions, the moment equations obtained from the master equation cannot generally be solved in closed form, and thus various approximations of the master equation have been developed. In particular for systems composed of unimolecular and bimolecular reactions, the time-evolution equation of the $M^{th}$ central moment of the probability density function solution of the master equation is generally a function of the $(M+1)^{th}$ central moment. This implies an infinite hierarchy of coupled equations which cannot be generally solved \cite{Uribe2007}, although see \cite{Stefanini05} for an exception. In the limit of intermediate or large numbers of molecules, various methods have been developed to obtain approximate expressions for the moments (examples of two widely used methods are the system-size expansion \cite{vankampen_1992,ElfEhrenberg,Newman2007,Grima2009,Grima2010,Grima2011} and moment-closure approximations \cite{Uribe2007,Goutsias2007,Ferm2008,Ullah2009,Grima2012}). However the reliability and accuracy of these methods when applied to systems characterized by low copy number of molecules and bimolecular reaction steps has remained an outstanding question of practical interest. Hence we anticipate that our exact solution will also provide a useful benchmark with which to compare the gamut of approximation methods used to estimate the effect of noise in biochemical systems.
 
\acknowledgments 
Two of the authors (DRS and TJN) gratefully acknowledge the 2007-08 program on Mathematics of Molecular and Cellular Biology, organised by the Institute for Mathematics and its Applications, University of Minnesota, where this project was initiated. RG and TJN acknowledge support from SULSA (Scottish Universities Life Sciences Alliance). DRS acknowledges support from NSF agreement 0112050 through the Mathematical Biosciences Institute.

\begin{flushleft}

\end{flushleft}
 
\appendix 

\section{The case R=0}
 
The exact solution presented in Eqs. (\ref{w_defn}), (\ref{alpha_defn}), (\ref{beta_defn}) and (\ref{gf1_exact}) assumes that the parameter $R=\rho _{u}-\rho_{b} \Sigma _{b}$ is different to zero. The case $R=0$ requires a separate analysis, either by taking the Kummer function solution (\ref{gf1_exact}) and taking a careful limit, or else by reexamining the second-order differential equation (\ref{gf1_ode}) explicitly for $R=0$. For illustrative purposes, we demonstrate the second method here, and derive an exact expression for $G_{1}(z)$ and $P_{1}(n)$. These calculations can be extended to derive the exact form for $P_{0}(n)$, if desired by the reader. 

Starting with Eq.(\ref{gf1_ode}) we set $R=0$ by writing $\rho_{b}=\rho_{u}/\Sigma_{b}$. We make the double transformation
\begin{equation}
\label{gf1_tilde_ap1}
G_{1}(z)=\exp (\rho_{u}/\Sigma _{b}) \ {\tilde G}_{1}(z) \ ,
\end{equation}
and
\begin{equation}
\label{u_fn_z}
u=(z-1/\Sigma _{b})^{1/2} \ , 
\end{equation}
to obtain the differential equation for ${\tilde G}_{1}(u)$ which has the form
\begin{equation}
\label{gf1_tilde_ode}
\frac {d^{2}{\tilde G_{1}}} {du^{2}} + \frac{(2\gamma + 1)}{u}\frac {d{\tilde G_{1}}} {du} - \lambda {\tilde G}_{1} = 0  \ ,
\end{equation}
where
\begin{equation}
\label{gamma_defn}
\gamma = \theta + \frac {\sigma _{u}}{\Sigma _{b}} + \frac {\rho_{u}(\Sigma _{b}-1)}{\Sigma _{b}^{2}}  \ , 
\end{equation}
and
\begin{equation}
\label{lambda_defn}
\lambda = \frac {4\rho_{u}\sigma _{u}(\Sigma _{b}-1)}{\Sigma _{b}^{2}} \ .
\end{equation}
Eq. (\ref{gf1_tilde_ode}) can be directly related to the differential equation for the Bessel function \cite{abramowitz_1972}, and we have
\begin{equation}
\label{gf1_exact_ap}
{\tilde G}_{1}(z) = A u^{-\gamma} I _{\gamma }(\lambda ^{1/2} u) \ ,
\end{equation}
where $A$ is a normalisation constant and $I_{\gamma}$ is the modified Bessel function. Note, $I_{-\gamma}$ is an independent solution (for $\gamma $ not equal to an integer), but can be discarded, since its asymptotic properties lead to a non-normalisable probability distribution.

In order to derive an explicit form for $P_{1}(n)$ we use Eq. (\ref{invert_gf}), and the differentiation formula \cite{abramowitz_1972}
\begin{equation}
\label{bessel_diff}
\left ( \frac {1}{v} \frac {d}{dv} \right )^{n} v^{-\gamma}I_{\gamma }(v) = v^{-\gamma -n}I_{\gamma +n}(v)  \ .
\end{equation}
This provides the final result
\begin{equation}
\label{bessel_diff}
P_{1}(n) = \frac {A'}{n!} \sum \limits _{m=0}^{n} C_{m}^{n} 
\left ( \frac {\rho_{u}}{\Sigma _{b}}  \right ) ^{n-m}
\left ( \frac {\lambda}{2}  \right )^{m} 
\left ( \frac {\Sigma _{b}}{\lambda} \right )^{(m+\gamma)/2}
J_{\gamma +m} \left ( \left ( \lambda/\Sigma _{b}  \right )^{1/2} \right )   \ ,
\end{equation}
where $A'$ is a normalisation constant. The reason the Bessel function $J$ appears in $P_{1}$ rather than the modified Bessel function $I$, is due to the argument of Eq. (\ref{gf1_exact_ap}) becoming imaginary when $z=0$. 
 
\section{Mean-field theory} 
  
We consider a fictitious experiment in which the cell membranes of a large number of identical cells enclosed in some reaction volume $\Omega$ are dissolved such that the genes and proteins from each individual cell can interact with that from every other cell. The well-mixed dynamics of this reaction system is deterministic (due to the large number of cells) and the state of the system at any point in time is described by two variables: the concentration of unbound DNA molecules, $\phi_u$,  and of the protein, $\phi$. Note that the concentration of bound DNA molecules is $\phi_b = \phi_T - \phi_u$ (where $\phi_T$ is the concentration of bound and unbound DNA) and is hence not an independent variable. The conventional mass-action rate equations for the concentrations can be directly deduced by inspection of the reaction scheme Eq. (\ref{scheme}) and are given by
\begin{align}
\partial_t \phi_u &= k_b (\phi_T - \phi_u) - k \phi_u  \phi  + s_u (\phi_T - \phi_u), \\
\partial_t \phi &= r_u \phi_u  + r_b (\phi_T - \phi_u) - k_f \phi - k \phi_u  \phi + s_u (\phi_T - \phi_u).
\end{align}
Note that the bimolecular rate constant $k$ is not $s_b$ but rather is equal to $s_b \Omega$. This is since $s_b$ is a transition rate with units of inverse time and hence is in reality equal to the macroscopic rate constant $k$ (with units of volume divided by time) divided by the reaction volume. Note also that these equations are based on the implicit assumption that the covariance of fluctuations in the number of molecules of any pair of species is zero, i.e. fluctuations are not important.

In order to compare with the single gene results derived in the main text, we first multiply the above equations by the reaction volume $\Omega$ and then set this volume equal to the cellular volume, i.e. $\Omega \phi_T = 1$, which leads to mean-field equations for the average molecule numbers of unbound DNA $\langle n_u \rangle$ and of protein $\langle n \rangle$ in a cell:
\begin{align}
\label{MFT1eqA}
\partial_t \langle n_u \rangle &= k_b (1 - \langle n_u \rangle) - s_b \langle n_u \rangle \langle n \rangle + s_u(1 - \langle n_u \rangle), \\
\label{MFT2eqA}
\partial_t \langle n \rangle &= r_u \langle n_u \rangle + r_b (1 - \langle n_u \rangle) - k_f \langle n \rangle - s_b \langle n_u \rangle \langle n \rangle + s_u (1 - \langle n_u \rangle).
\end{align}
These equations admit a single steady-state solution for $\langle n_u \rangle$ and $\langle n \rangle$ implying that the deterministic model does not predict bistability. 

Within this approach, the quantity $1 - \langle n_u \rangle$ can be interpreted as the fraction of time that the promoter is bound, $f_{\rm off}$. Substituting in the latter equation, the steady-state solution of Eq. (\ref{MFT1eqA}) for $\langle n_u \rangle$ and using the dimensionless rates as given by Eq. (\ref{drates}), one obtains the Hill equation
\begin{equation}
f_{\rm off} = \frac{\langle n \rangle}{\langle n \rangle + \frac{\theta + \sigma_u}{\sigma_b} }.
\end{equation}

\newpage

\begin{figure}[ht]
\centering
\subfigure[]{
\includegraphics[width=2.5in]{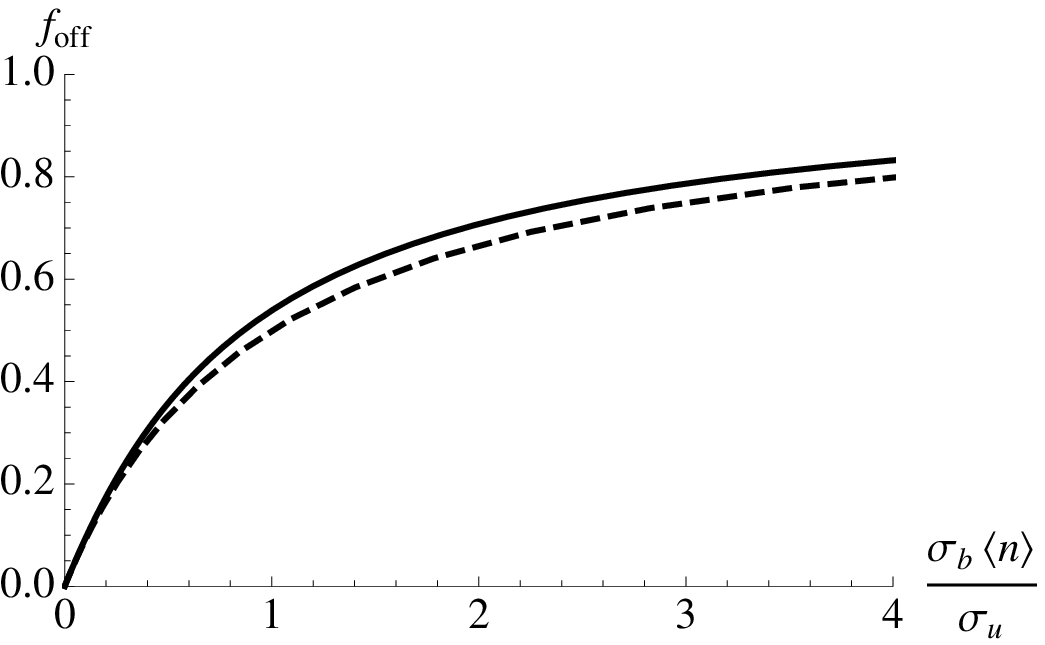}
\label{fig:subfig1}
}
\subfigure[]{
\includegraphics[width=2.5in]{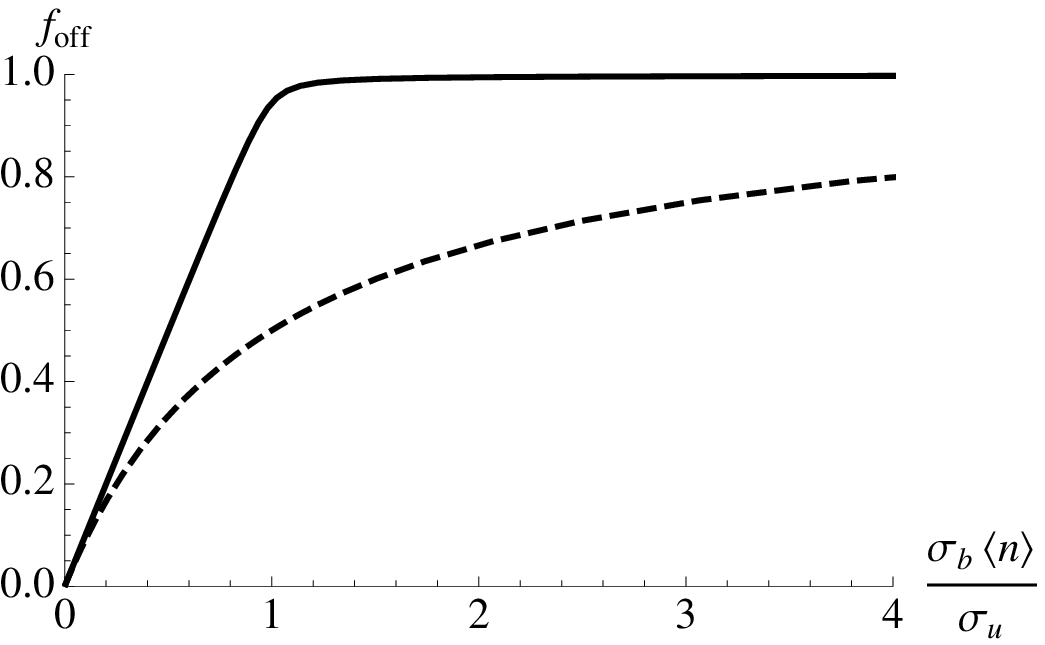}
\label{fig:subfig2}
}
\caption[]{Plot of $f_{\rm off}$ as a function of $\langle n \rangle$ for the case $\rho_b = 0$. The solid lines are generated by evaluating Eqs. (\ref{frac_off}) and (\ref{av_n}) for varying values of $\rho_u$. We also plot, for comparison, the Hill function (dashed lines) given by Eq. (\ref{hillapprox}), whose small and large $\langle n \rangle$ limits agree with those of the actual function (the one given by solid lines). Note that the Hill function is also the prediction of the deterministic model of the genetic feedback loop (Appendix B). The parameters are $\sigma_b = 0.01, \sigma_u = 2$ in (a) and $\sigma_b = 2, \sigma_u = 0.01$ in (b). In both cases $\theta = 0$. Note that the Hill function approximates well the actual function for small $\sigma_b$; this is generally the case for all $\theta$.}
\end{figure}
  
\newpage

\begin{figure}[ht]
\centering
\subfigure[]{
\includegraphics[width=2.5in]{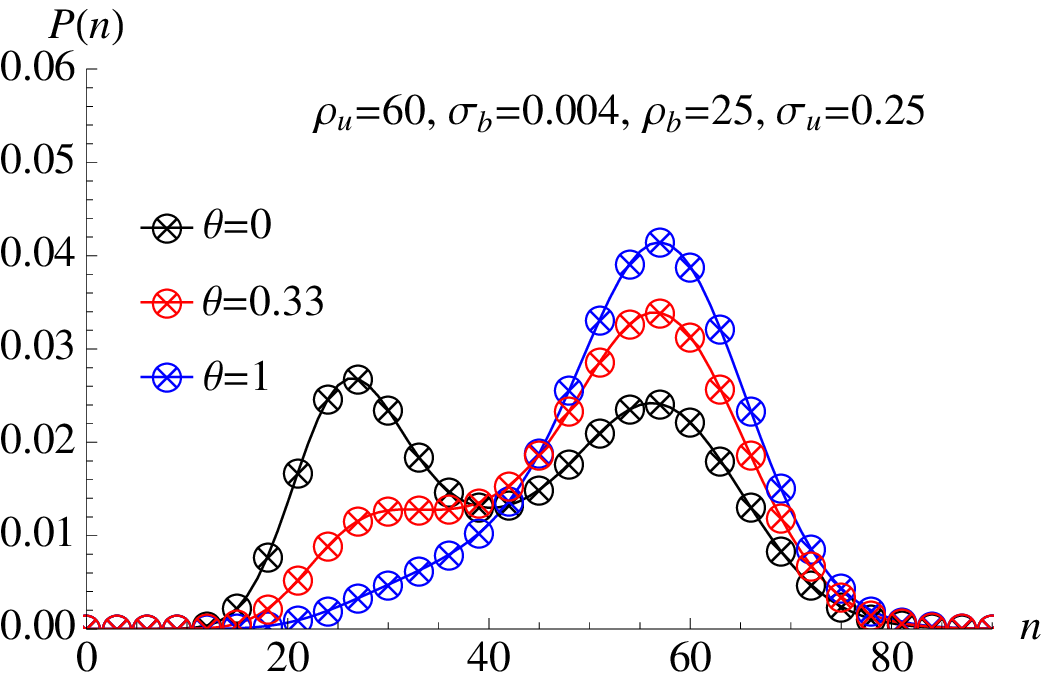}
\label{fig:subfig1}
}
\subfigure[]{
\includegraphics[width=2.5in]{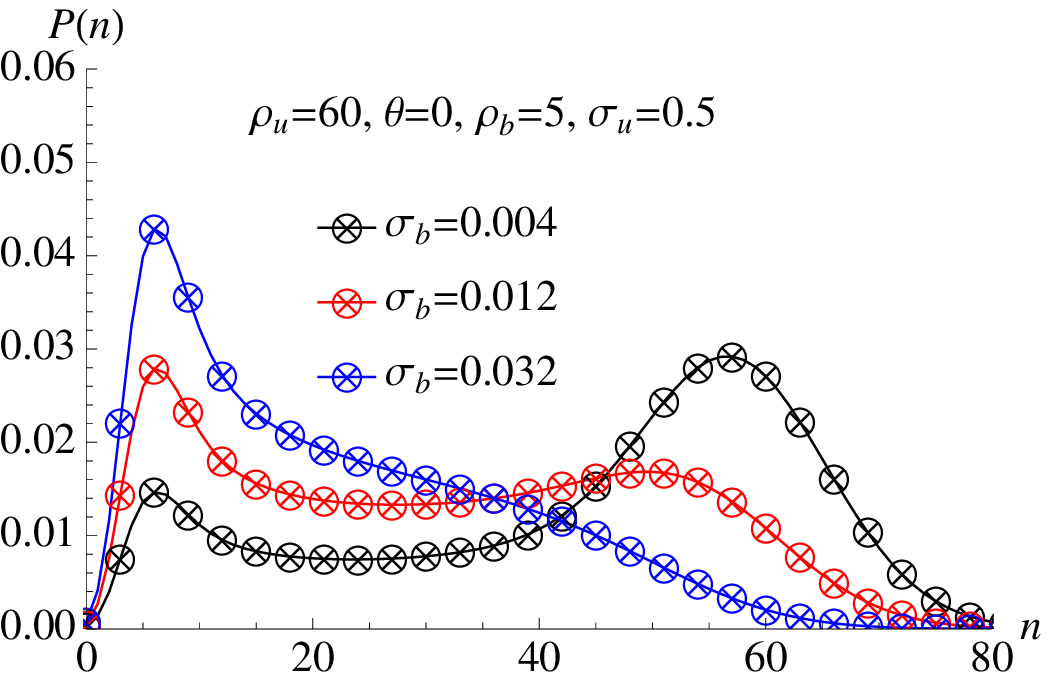}
\label{fig:subfig2}
}
\subfigure[]{
\includegraphics[width=2.5in]{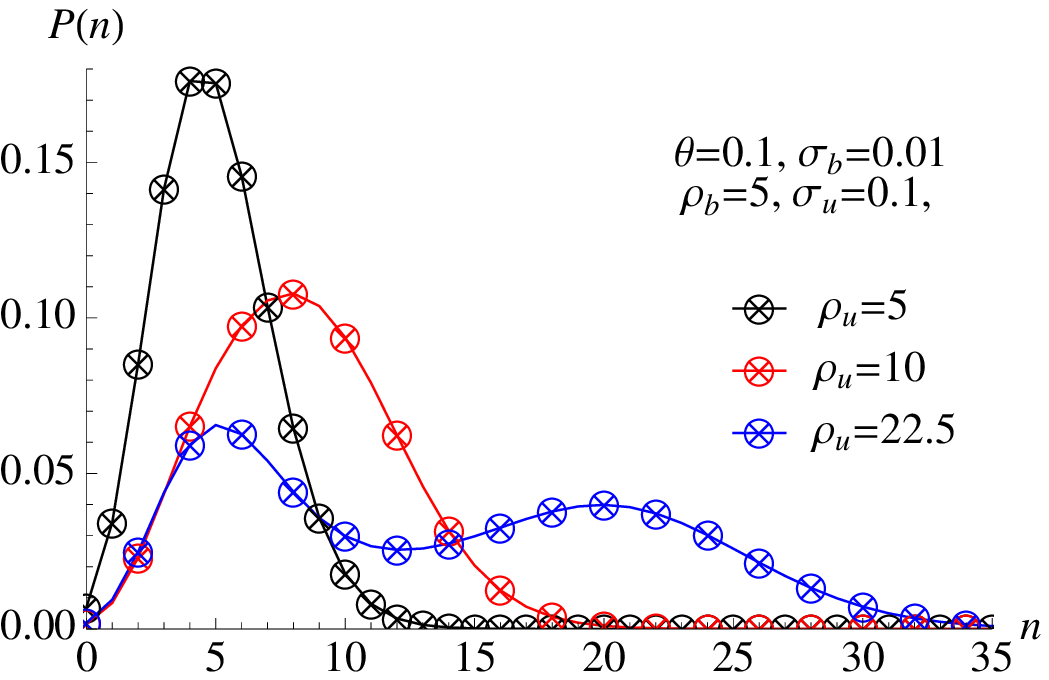}
\label{fig:subfig1}
}
\subfigure[]{
\includegraphics[width=2.5in]{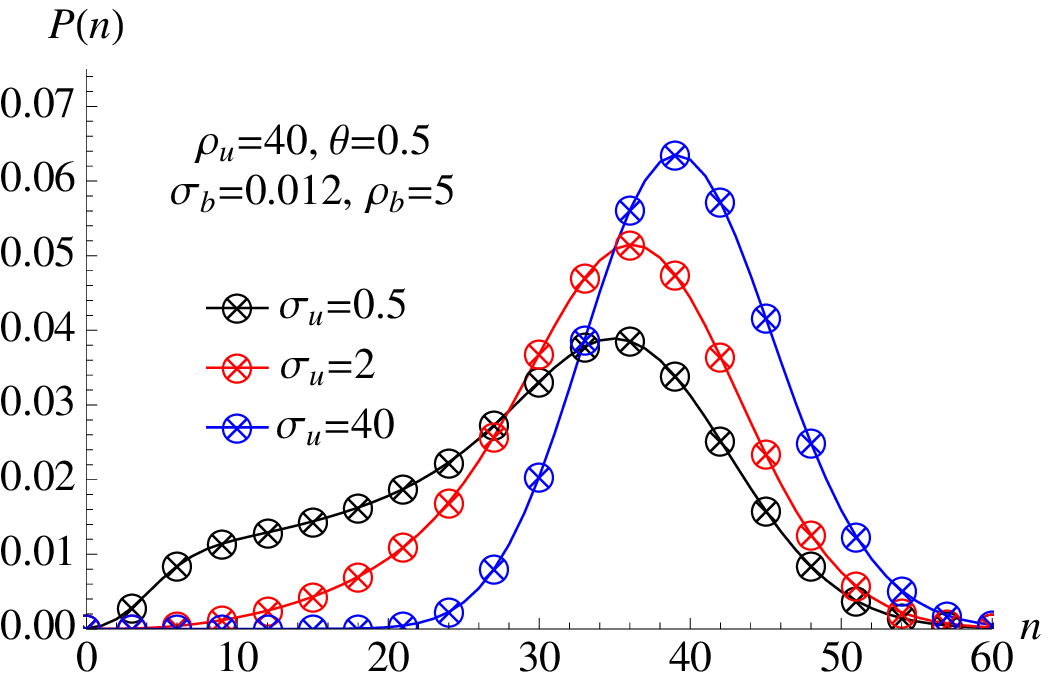}
\label{fig:subfig2}
}
\subfigure[]{
\includegraphics[width=2.5in]{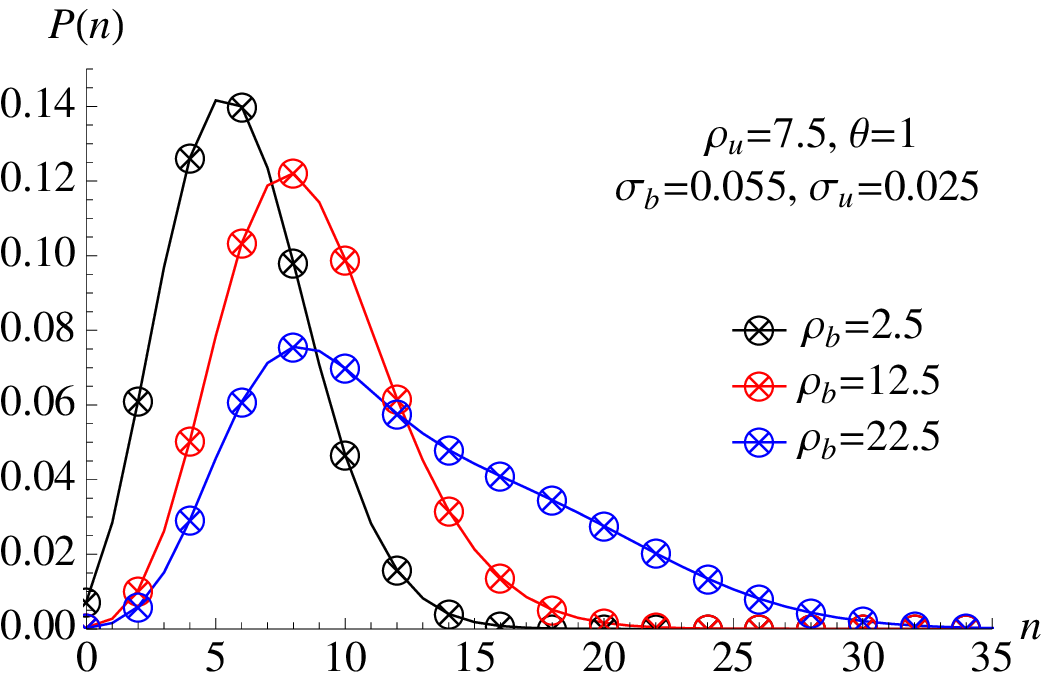}
\label{fig:subfig1}
}
\subfigure[]{
\includegraphics[width=2.5in]{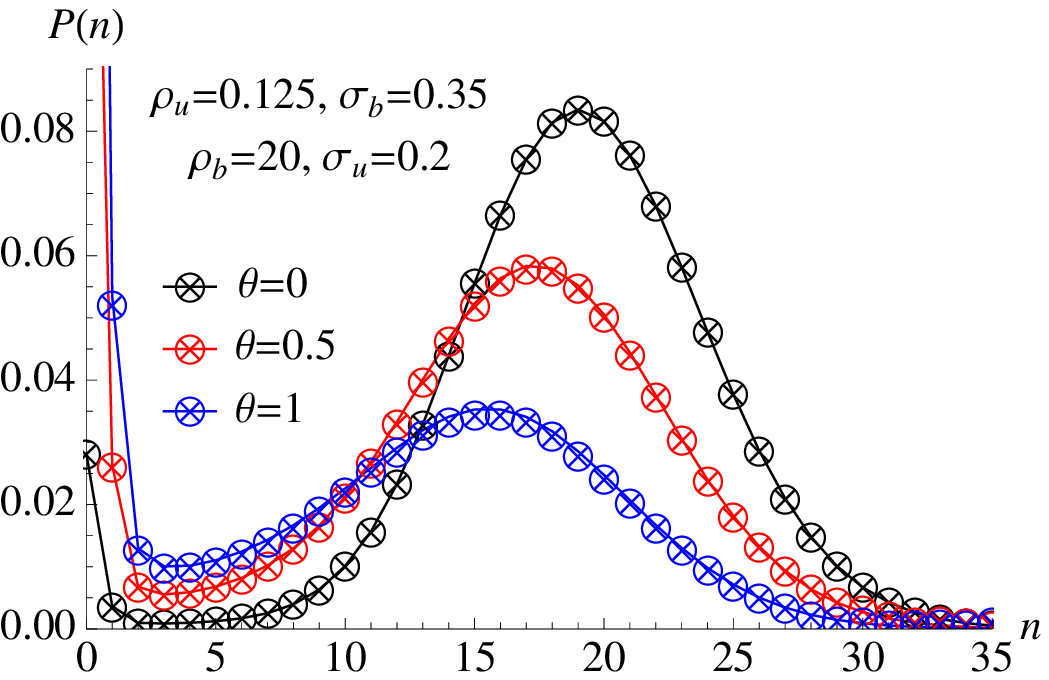}
\label{fig:subfig1}
}
\caption[]{Dependence of the steady-state probability distribution on  
the non-dimensional parameters $\theta$ (panel a and f), $\sigma_b$  
(panel b), $\rho_u$ (panel c), $\sigma_u$ (panel d), and $\rho_b$  
(panel e). The distributions are obtained by numerically integrating  
the master equations, Eqs. (\ref{master_dim_1}) and (\ref{master_dim_2}) (shown by  
the open circles), and by evaluating the analytical solutions, Eqs.  
(\ref{p1_exact}), (\ref{p0_exact}) and (\ref{p0_0_exact}) (shown by the  
crosses). The perfect agreement of the two  
verifies that the latter equations are an exact solution of the master  
equation for the self-regulating gene. The solid lines are a guide to the eye.
}
\end{figure}

\newpage 
 
\begin{figure}[ht]
\centering
\subfigure[]{
\includegraphics[width=2.5in]{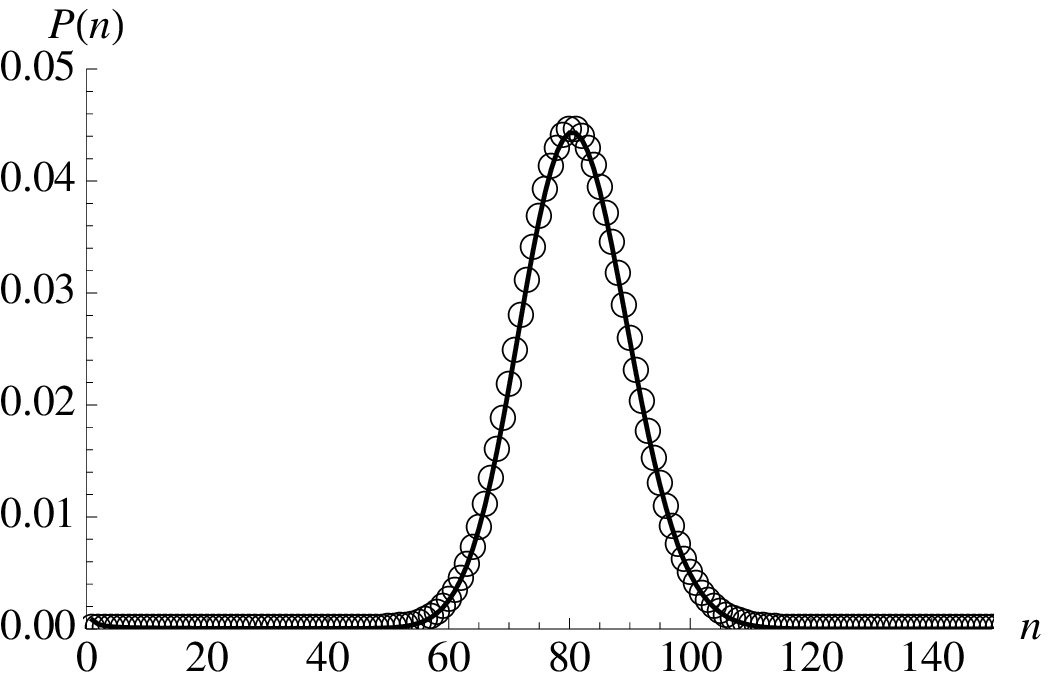}
\label{fig:subfig1}
}
\subfigure[]{
\includegraphics[width=2.5in]{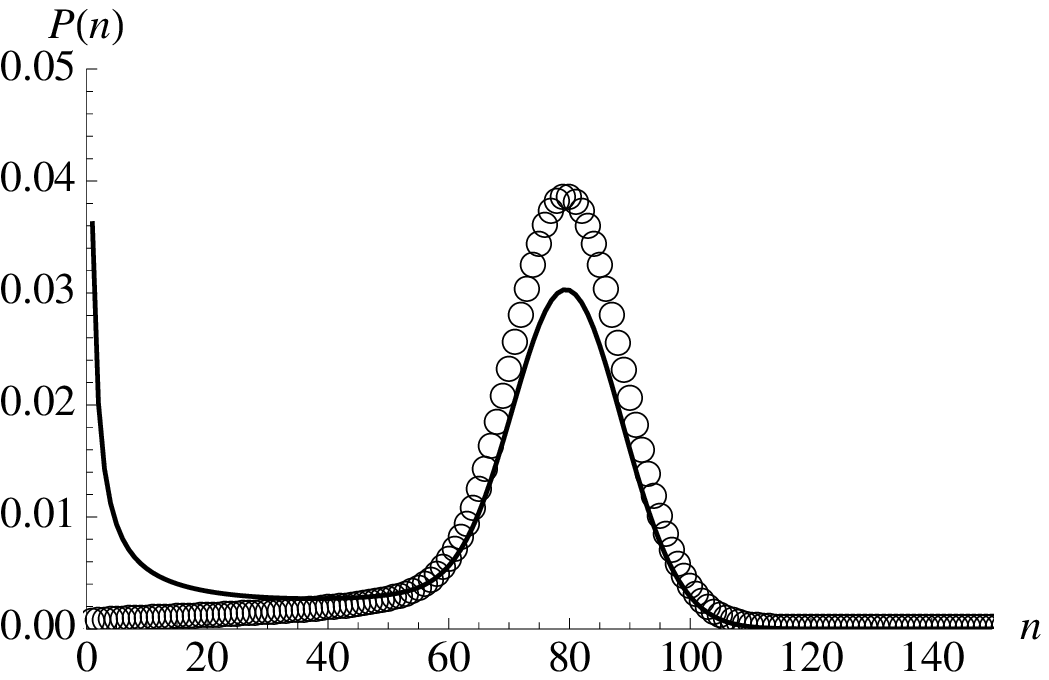}
\label{fig:subfig2}
}
\subfigure[]{
\includegraphics[width=2.5in]{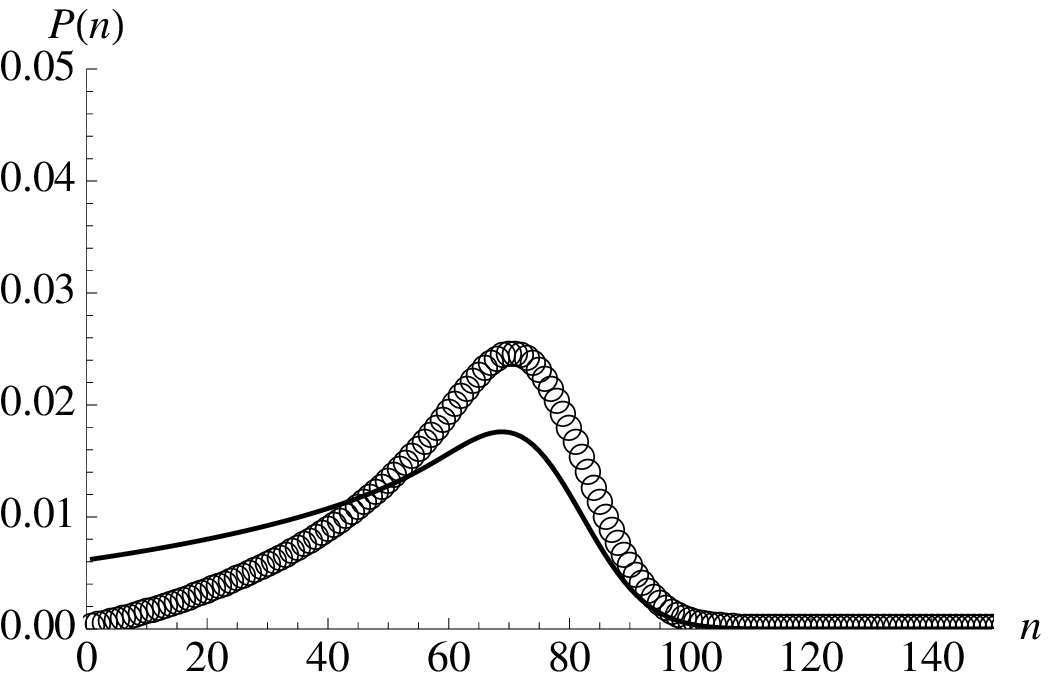}
\label{fig:subfig1}
}
\subfigure[]{
\includegraphics[width=2.5in]{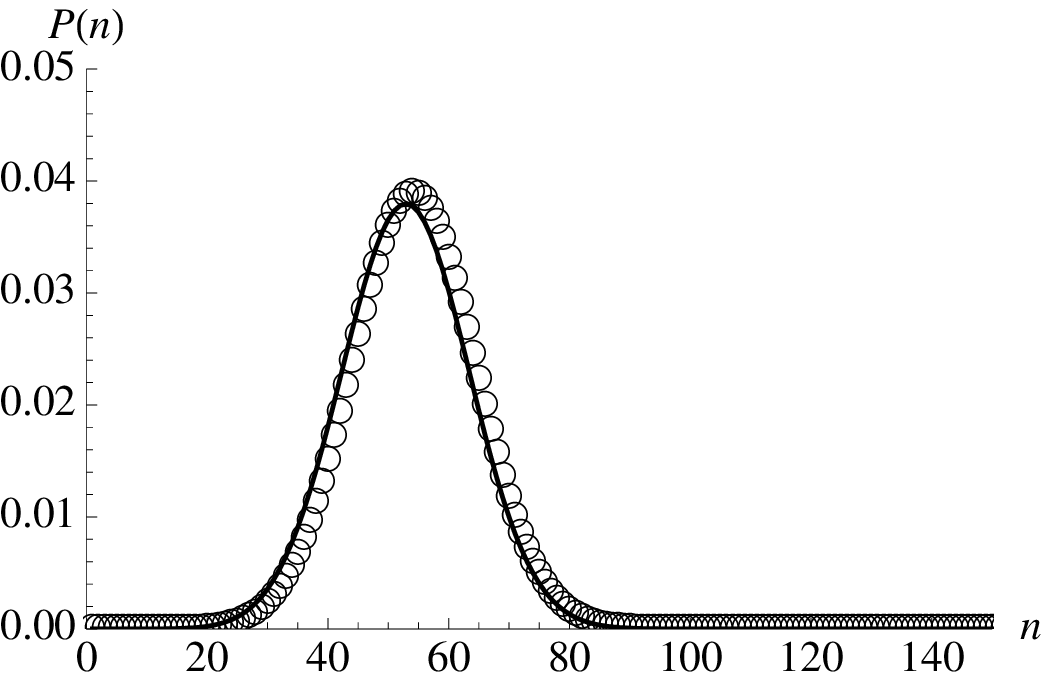}
\label{fig:subfig2}
}

\caption[]{Comparison of the steady-state probability distributions as predicted by the master equations ME(GSN) (open circles) and ME(H-ET-AL) (solid lines) . The distributions are obtained by numerically integrating the two master equations to obtain $P(n) = \alpha_n+\beta_{n+1}$ as a function of $n$, the number of free proteins. The binding and unbinding rates of the protein to the promoter region, $h$ and $f$ respectively, take progressively larger values as we go from (a) to (d), while the rest of the parameters are fixed. Explicitly, $g_{\alpha}$, $g_{\beta}$, and $k$ are, in all cases, equal to $80.0$, $0.0$, and $1.0$ respectively, while $h = 0.001\Lambda$ and $f = 0.1\Lambda$ where $\Lambda$ takes the values 0.01 in (a), 1 in (b), 10 in (c) and 100 in (d). Note that the ME(H-ET-AL) predicts a transition from unimodal (a) to bimodal (b) and back to unimodal probability distribution (c) and (d), while the ME(GSN) predicts a unimodal distribution in all cases. This example shows that the incorrect handling of bound protein degradation by the ME(H-ET-AL) leads to qualitatively incorrect features of the steady-state probability distribution. 
}
\end{figure}
 

\begin{thebibliography} {10}
\bibitem{Cmt} The principle of detailed balance asserts that for any pair of states $n$ and $n'$, the probability of being in state $n$ multiplied by the probability per unit time of jumping from $n$ to $n'$ is equal to the probability of being in state $n'$ multiplied by the probability per unit time of jumping from $n'$ to $n$. Hence detailed balance implies that pairs of terms in the master equation are separately zero whereas steady-state conditions require that the whole sum of terms is zero. The stronger condition of local equilibrium imposed by detailed balance leads to an exact solution of the master equation.
\end{thebibliography}
\end{document}